\documentclass[12pt,letterpaper]{JHEP3}
\usepackage{cite}
\usepackage{epsfig}
\usepackage{graphicx}
\usepackage{subfigure}
\usepackage{amsmath}
\usepackage{amssymb}
\graphicspath{{Figures/}}

\title{Bubble Nucleation of Spatial Vector Fields}

\author{Ali Masoumi\footnote{ali@phys.columbia.edu}~, 
Xiao Xiao\footnote{xx2146@columbia.edu}~, and 
I-Sheng Yang\footnote{isheng.yang@gmail.com} \\
\em ISCAP and Physics Department \\
Columbia University, New York, NY, 10027 , U.S.A. \\
}

\abstract{We study domain-walls and bubble nucleation in a non-relativistic vector field theory with different longitudinal and transverse speeds of sound. We describe analytical and numerical methods to calculate the orientation dependent domain-wall tension, $\sigma(\theta)$.  We then use this tension to calculate the critical bubble shape.  The longitudinally oriented domain-wall tends to be the heaviest, and sometime suffers an instability.  It can spontaneously break into zigzag segments.  In this case, the critical bubble develops kinks, and its energy, and therefore the tunneling rate, scales with the sound speeds very differently than what would be  expected for a smooth bubble.}

\begin{document}

\section{Introduction and Outline}

The study of first-order phase transitions is a fascinating subject that appears in many branches of physics.  The standard picture is to nucleate (thermally or quantum mechanically) a bubble in a homogeneous background of the false vacuum. The bubble interior is in the true vacuum, and it is surrounded by domain-walls---the minimal energy field interpolation between the false and true vacua.

In this paper we will focus on thermal nucleation, where the critical bubble is the lowest saddle point of the energy barrier. For thermal tunneling, the time variable is not that important and focusing on theories which have vectors transforming under the spatial rotation can shed light on many of the subtleties.  In the simplest example, a scalar field theory, one can show that the critical bubble must have $SO(N)$ symmetry in $N$ dimensional space\cite{ColGla77,Col87}.  This leads to the commonly used estimate for the tunneling rate $\Gamma$, 
\begin{eqnarray}
\log \Gamma &\sim& -\frac{E_s}{k_b T}
\sim \frac{\sigma^N}{\Delta V^{N-1}}\frac{1}{k_b T}~, \\
\sigma &=& v_F \int_{\text{path in field space}} \!\!\!\!\!\!\!\!\!\!\!\!\!\!\!\!\! \!\!\!\!\!\! \!\!\!\!\!\!  d\,\phi \: \;\,\, \,\sqrt{2V}~.
\label{eq-rate}
\end{eqnarray}
Here the critical bubble energy $E_s$ is determined by the domain-wall tension $\sigma$ and the energy difference $\Delta V$, assuming a spherical bubble.  The tension is given by the path in the field space which minimizes that integral. The scaling with the Fermi velocity $v_F$ follows from the equation of motion.

In this paper we will generalize the theory to include vector fields. Our motivation comes form condensed matter systems like liquid crystals, Helium 3 and Langmuir monolayers\cite{Leg75,Whe75,He3,GalFou95,Fou95,MacJia95,RudLoh99,SilPat06}.  The simplest vector fields to imagine are the non-relativistic vector fields transforming under the spatial rotation group. In $(n+1)$-dimensional spacetime, these vectors have $n$ components. We study the transitions between two discrete minima $\vec{\phi}_\pm$ of a field with the Lagrangian
\begin{equation}
\mathcal{L}=\frac{1}{2}
\left(\dot{\phi}_i^2 - c_T^2 \partial_i\phi_j\partial_i\phi_j - (c_L^2-c_T^2) \partial_i\phi_i\partial_j\phi_j \right) - V(\phi_i)~.
\label{eq-L}
\end{equation}
In order to make the energy bounded from below, we need $c_L \geq c_T$ . When $c_L\neq c_T$, the potential can minimally break  the spatial $SO(N)$ symmetry.  We will focus on the case with minimal breaking.  For any field configuration that involves these two vacua, at least $(\vec\phi_+-\vec\phi_-)$ is a special direction that specifies the longitudinal wall and breaks the symmetry down to $SO(N-1)$.

In Sec.\ref{sec-orient}, we study planar domain-walls.  Due to the broken symmetry, the domain-wall tension acquires an orientation dependence.  We set up the general analytical and numerical process to determine $\sigma(\theta)$, where $\theta$ is defined as the angle between the normal vector of the wall and $(\vec\phi_+-\vec\phi_-)$.  We demonstrate a rich behavior of $\sigma(\theta)$ through examples in Appendix.\ref{sec-examples}.  We further show that in the orientations which the domain-wall is heavy, it may develop an instability and spontaneously break into zigzag segments of lighter walls.

In Sec.\ref{sec-shape}, we solve for the shapes of critical bubbles from $\sigma(\theta)$.  The solution has a simple form when the above stated instability does not occur.  When it does, the function describing the bubble shape becomes multi-valued.  We show that it still has a simple interpretation and describes bubbles with kinks.  We then calculate how the deformed critical bubble modifies the transition rate.

The technique we use to solve for the shapes is identical to that for equilibrium bubbles, known as the Wulff construction\cite{RudBru95}.  It has been applied to ``soft matter'' systems like liquid crystals and Langmuir monolayers\cite{GalFou95,Fou95,MacJia95,RudLoh99,SilPat06}.  Our result agrees with the major conclusions in the these earlier works.  In Sec.\ref{sec-dis} we will summarize a few concepts sharpened by our analysis, and also provide an intuitive understanding of when and how the tunneling rates are modified.   

\section{Orientation Dependence}
\label{sec-orient}

The Lagrangian in Eq.~(\ref{eq-L}) leads to the following equation of motion,
\begin{equation}
\ddot{\phi}_i - c_T^2\partial_j^2\phi_i
- (c_L^2-c_T^2) \partial_i \partial_j\phi_j = 
-\frac{\partial V}{\partial \phi_i}~,
\label{eq-eom}
\end{equation}
where it is more apparent that $c_T$ and $c_L$ correspond to the transverse and the longitudinal sound speeds.

We want to have two isolated vacua in $V$.  This is quite easy to achieve using the following potential.
\begin{equation}
V(\vec{\phi}) = \frac{m^2}{2}|\vec{\phi}|^2 + 
\frac{\lambda}{4} |\vec{\phi}|^4 
+ a (\vec{H}\cdot\vec{\phi}) + b (\vec{H}\cdot\vec{\phi})^2~.
\label{eq-potmot}
\end{equation}
The last two terms are the two lowest orders of the effect from an external field $\vec{H}$.  We start by considering $a=0$, then $b<0$ picks a preferred direction along $\vec{H}$.  When $b|\vec{H}|^2 + m^2/2 <0$, we get two degenerate vacua at 
\begin{equation}
\vec{\phi}_\pm = \pm\sqrt{\frac{m^2+2b\vec{H}^2}{\lambda}}\frac{\vec{H}}{|\vec{H}|}~.
\end{equation}
Afterward, a small $a$ can break the degeneracy to allow a first-order phase transition.  This is just an example to show how achievable our setup is.  Our further analysis will either be independent of the form of the potential, or focus on examples similar but even simpler than Eq.~(\ref{eq-potmot}).

To study first-order phase transitions, a useful starting point is the thin-wall bubble.  First we pretend that the two vacua are degenerate and find an interpolation between them, which is a domain-wall.  The property of the domain-wall will then later be used to form a bubble of the nucleation event.  

\begin{figure}
   \centering
   \includegraphics[width=12cm]{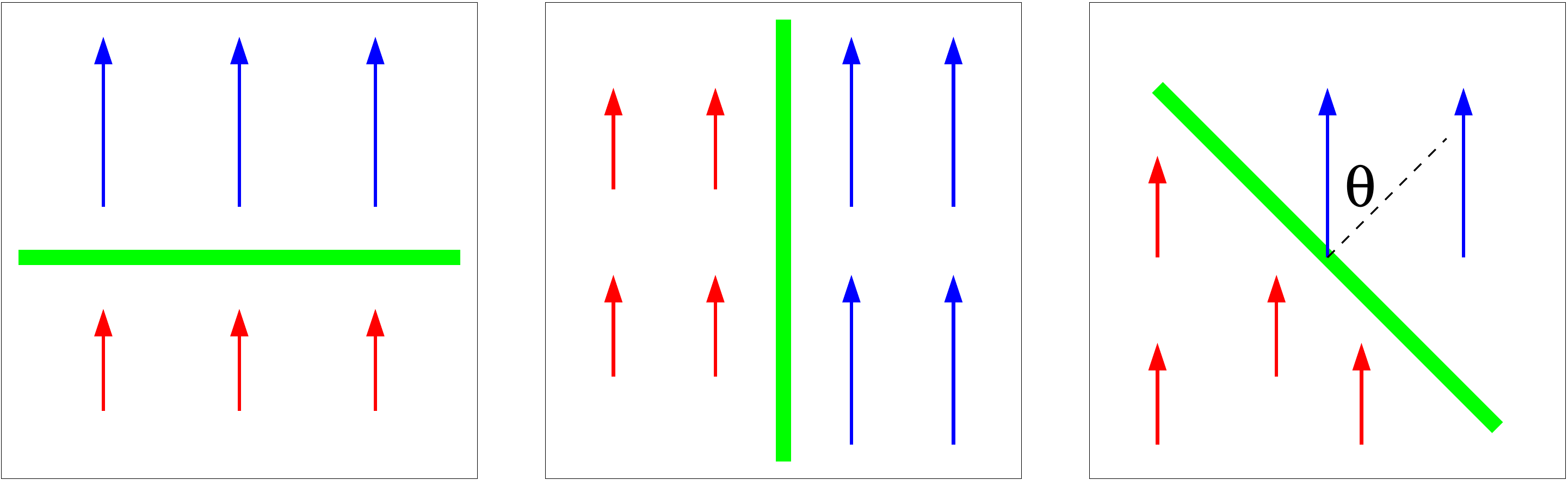} 
   \caption{The blue (longer) and red (shorter) arrows represent the vector field value of two vacua.  The thick green line is the domain-wall.  From left to right, we show a longitudinal wall, a transverse wall, and a wall with orientation $\theta$.  The orientation is defined such that for a longitudinal wall $\theta=0$, and for a transverse wall $\theta=\pi/2$.
\label{fig-orientation}}
\end{figure}

The important property already shows up when we consider the domain-wall.  Since the interpolation between the two vacua is a vector in the field space, it breaks the spatial rotational symmetry, as shown in Fig.~\ref{fig-orientation}.  How the vector $\vec\phi_+$ continuously changes into $\vec\phi_-$ can be a complicated process and clearly depends on the orientation.  In the thin wall approximation, we can summarize the effect as an orientation-dependent tension $\sigma(\theta)$.  When the tension is a constant, a first-order phase transition involves the nucleation of a spherically symmetric bubble.  So naturally in a vector field system, orientation dependence of $\sigma(\theta)$ can lead to a nontrivial bubble shape.  Here we will provide the general formalism to find $\sigma(\theta)$, and then in Sec.\ref{sec-shape} we will use it to find the bubble shape.

\subsection{Two Dimensions}

We will demonstrate our technique in the simplest example---a vector field in 2D.  For a potential with two degenerate vacua $\vec{\phi}_{\pm}$, a domain-wall is a static solution to the equation of motion,
\begin{eqnarray}
-c_T^2(\partial_x^2+\partial_y^2)\phi_x
-(c_L^2-c_T^2)\partial_x(\partial_x\phi_x+\partial_y\phi_y)
&=&-\frac{\partial V}{\partial\phi_x}~,\nonumber \\
-c_T^2(\partial_x^2+\partial_y^2)\phi_y
-(c_L^2-c_T^2)\partial_y(\partial_x\phi_x+\partial_y\phi_y)
&=&-\frac{\partial V}{\partial\phi_y}~.
\end{eqnarray}
The boundary condition is specified by two orthogonal vectors $\vec{u}\cdot\vec{v}=0$, such that \begin{eqnarray}
(\vec{v}\cdot\nabla)\vec{\phi} &=& 0~, \label{eq-sym} \\
\lim_{\lambda\rightarrow\pm\infty}\vec{\phi}(\lambda\vec{u})
&=&\vec{\phi}_{\pm}~.
\end{eqnarray}
Namely, the field value interpolates between the two vacua purely along the normal vector of the domain-wall, $\vec{u}$.

For simplicity, we can actually always choose $\vec{u}=\vec{y}$ and instead apply a rotation on the potential,
\begin{equation}
V_\theta(\phi_x,\phi_y)=V(\phi_x\cos\theta+\phi_y\sin\theta,\phi_y\cos\theta-\phi_x\sin\theta)~.
\end{equation}
This simplifies the equation of motion to
\begin{eqnarray}
- c_T^2 \partial_y^2\phi_x &=& 
-\frac{\partial V_\theta}{\partial\phi_x}~, \nonumber \\
- c_L^2 \partial_y^2\phi_y &=& 
-\frac{\partial V_\theta}{\partial\phi_y}~.
\label{eq-eom2}
\end{eqnarray}
The solution we get here is the $\vec{y}$ oriented domain-wall in potential $V_\theta$, which is equivalent to the $\vec{u}$ oriented domain-wall in the original potential $V$ with $\hat{u}\cdot\hat{y}=\cos\theta$.

The tension of the domain-wall is given by the total energy per unit $x$.
\begin{eqnarray}
\sigma(\theta) = \int dy \left[ \frac{1}{2}\left(c_L^2\phi_y'^2+c_T^2\phi_x'^2\right)+V_\theta \right]~.
\label{eq-tension}
\end{eqnarray}
Here we set $V=0$ in the vacua.  It is well-known that the practical way to find the domain-wall solution is to numerically minimize this tension\cite{AguJoh09a,GibLam10,AhlGre10}, which is what we do in Appendix~\ref{sec-examples}.

In principle, the orientation dependence of $\sigma$ can be arbitrarily complicated through $V_\theta$. Here we would like to start from a simple, yet in some sense typical case.  Imagine the situation where at $\theta=0$, the interpolation is purely longitudinal, $\phi_x=const$. \footnote{Note that we talk about a particular solution, instead of imposing some symmetry on $V$.  This is necessary.  One might try a rotational (reflection in the 2D case) symmetry on $V$ along the vector $(\vec\phi_+-\vec\phi_-)$.  That turns out to be not necessary nor sufficient to guarantee that $\phi_x$ is constant.}  Since a rotation of $\pi/2$ just exchanges $\phi_x$ and $\phi_y$, the interpolation will become purely transverse with $\phi_y=const$.  It is then easy to work out from Eq.~(\ref{eq-tension}) that
\begin{eqnarray}
\sigma(0) &=& c_L \int_{\rm path} \sqrt{2V} |d\vec\phi|~,\nonumber \\
\sigma(\frac{\pi}{2}) &=& c_T \int_{\rm path} \sqrt{2V} |d\vec\phi|~,
\label{eq-simpletension}
\end{eqnarray}
where the two integration paths are the same, so
\begin{equation}
\frac{\sigma(0)}{\sigma(\frac{\pi}{2})}=\frac{c_L}{c_T}~.
\label{eq-ratio}
\end{equation}

Potentials given by Eq.~(\ref{eq-potmot}) when $m^2>0$ satisfy the above assumptions, so do the simpler potentials we use in Appendix~\ref{sec-examples}. They not only show a good agreement with Eq.~(\ref{eq-ratio}), but also demonstrate an excellent fit to a na\"ive interpolation,
\begin{equation}
\sigma(\theta) = \sigma(0)\cos^2\theta + \sigma(\frac{\pi}{2})\sin^2\theta~,
\label{eq-sigma}
\end{equation}
in the regular range of parameters.  From the symmetry of the problem, it seems natural to expand $\sigma(\theta)$ as a polynomial of $\sin^2\theta$ and keep the lowest order terms.

We also analyze two extreme choices of parameters in Appendix~\ref{sec-examples}. One of them corresponds to the following tension.
\begin{equation}
\sigma(\theta) = \sqrt{\sigma(0)^2\cos^2\theta+
\sigma(\frac{\pi}{2})^2\sin^2\theta}~.
\label{eq-sigmamotion}
\end{equation}

It turns out that Eq.~(\ref{eq-sigma}) and (\ref{eq-sigmamotion}) are quite representative for our further analysis.  Despite their simple forms which will simplify the calculation, they can actually have dramatically different behaviors.

\subsection{Flat Wall Instability}
\label{sec-kink}

Now we can think about a very practical question.  Since $c_L>c_T$, $\sigma(0)$ is most likely the maximum tension.  Even if the boundary condition is set up to preserve the $x$ translational symmetry,  the minimum energy interpolation can spontaneously break that symmetry.  In plain words, we might be able to replace the flat wall with a large tension $\sigma(0)$ by non-flat walls with smaller tensions, and hence reduce the total energy.

The full treatment of this problem is to remove condition~(\ref{eq-sym}) and see if a symmetry breaking configuration can further minimize the total energy. That is a quite involved numerical work which we will not pursue in this paper.
We will simply demonstrate this possibility in the thin-wall approximation.

The total energy is a functional of the domain-wall shape, $y(x)$.
\begin{equation} \label{eq-PureWall}
	E[y(x)] = \int_{x_1}^{x_2}
	 \sigma(-\tan^{-1} y') \sqrt{1+ y'^2} dx = 
	 \int_{x_1}^{x_2} \frac{\sigma(\theta)}{\cos\theta} dx ~.
\end{equation}
Now, given a symmetric boundary condition $y(x_1)=y(x_2)$ that na\"ively asks for a flat wall, we can ask two questions:
\begin{itemize}
\item Is the solution with $y'=-\tan\theta=0$ a stable minimum of the total energy?
\item Is there a different solution $y(x)$ that gives the global minimum?
\end{itemize}
In other words, is a flat wall perturbatively and non-perturbatively stable?

\begin{figure}
   \centering
   \includegraphics[width=10cm]{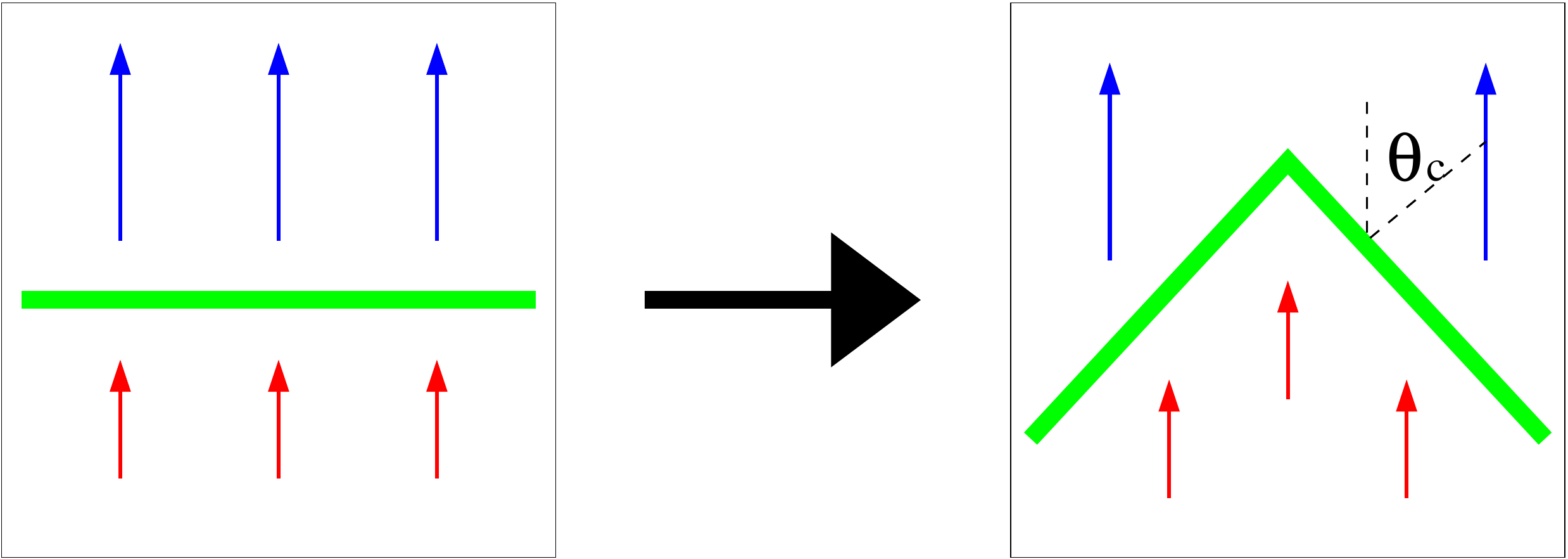} 
   \caption{When the flat, longitudinal wall on the left figure is too massive, it will spontaneously breaks into zigzag segments in the right figure.  Although the total wall area increases, the reduced tension still reduces total energy.
   \label{fig-zigzag}}
\end{figure}

Given that $\sigma(0)$ is a local maximum, expanding $E$ near $\theta=0$ gives
\begin{equation}
E \approx E_{\rm flat} + \int_{x_1}^{x_2} 
\left(\frac{1}{2}\frac{d^2\sigma}{d\theta^2}\bigg|_{\theta=0}  +\frac{\sigma(0)}{2}\right) \delta\theta^2 dx~.
\end{equation}
Thus, the perturbative stability condition is
\begin{equation}
\frac{1}{\sigma(0)}\frac{d^2\sigma}{d\theta^2}\bigg|_{\theta=0}>-1~.
\end{equation}
Next, the non-perturbative instability is about whether there is a $\theta\neq0$ such that
\begin{equation}
\sigma(\theta)<\sigma(0)\cos\theta~.
\end{equation}
When either or both instability exists, there will be a critical angle $\theta_c$ such  that $\sigma(\theta_c)/\cos\theta_c$ is the global minimum, and the wall prefers to settle into the zig-zag configuration that every segment is oriented at $\theta_c$, as shown in Fig.\ref{fig-zigzag}.  One can perform the same stability analysis for the other initial angles.  In this paper we will focus on simple cases where the global minimum $\theta_c$ is the only local minimum. It is straightforward to see that domain-walls more massive  than $\sigma(\theta_c)$ always break into zigzags, while the lighter walls are unaffected.

One may question the validity of our thin-wall analysis since usually the appearance of a kink means thick-wall effects are involved---it cannot be infinitely sharp and one needs to resolve the wall to understand it.  We should remind our readers that such concern is not important at this point.  Indeed a kink resolved by thick wall analysis will contribute a finite term to the total energy\footnote{This has been studied in various examples under the name ``boojum''.}, so it is not na\"ively energetic favorable to produce them.  However, in our planar wall setup, the $x$ and $y$ direction can be infinitely extended.  We only need a finite number of kinks to gain an arbitrarily large amount of energy by turning flat walls into $\theta_c$ zigzags, so our analysis is sufficient to determine whether it can happen.\footnote{This is just an example of the following general concept.  By definition, thin wall approximation works if the wall is thin---relatively to other length scales in the problem.  For a bubble it is compared to the bubble size.  In the flat wall setup, every other length scale is infinite, so the thin wall approximation is by definition good.}  When later talking about bubbles, we will see that only two kinks are necessary.  We can tune $\Delta V$ such that the bubble is arbitrarily large and approaches the flat wall situation, so the same logic applies.

Now take a look at the two examples for $\sigma(\theta)$, Eq.~(\ref{eq-sigma}) and (\ref{eq-sigmamotion}).  Their stability features are dramatically different.  If the orientation dependence is given by Eq.~(\ref{eq-sigmamotion}), then the flat wall is always stable.  On the other hand, the tension given by Eq.~(\ref{eq-sigma}) develops a perturbative instability once
\begin{equation}
\frac{\sigma(0)}{\sigma(\pi/2)}=\frac{c_L}{c_T}>2~.
\end{equation}
A flat wall with $\theta=0$ would break and settle into zigzag segments with 
\begin{equation}
\theta_c = 
\sin^{-1}\sqrt{\frac{\sigma(0) - 2 \sigma(\frac{\pi}{2})}
{\sigma(0) - \sigma(\frac{\pi}{2})}} 
=\sin^{-1} \sqrt{\frac{c_L- 2 c_T}{c_L- c_T}}~.
\end{equation}

\section{Bubble Shape}
\label{sec-shape}

Now we break the degeneracy between the two vacua by a small amount $\Delta V$ such that the thin wall approximation is still valid.  The phase transition mediated by a thermally nucleated bubble has the rate
\begin{equation}
\Gamma \sim \exp\left[-\frac{E_s}{k_bT}\right]~,
\end{equation}
where $E_s$ is the saddle point energy of the bubble.  We can find this saddle point by treating $E$ as a functional of the bubble shape $y(x)$,
\begin{eqnarray}
E[y(x)] &=& \sigma ({\rm surface \ area}) - \Delta V ({\rm volume})
\nonumber \\
&=& 4 \int ~dx~ \left[ \sigma\left(-\tan^{-1}y'\right) \sqrt{1+y'^2} 
- y ~ \Delta V \right],
\end{eqnarray}
where the symmetry allows us to cut the bubble into four quadrants.  We will focus on the first quadrant where $y'=-\tan\theta$. In order to perform functional variation, the standard boundary condition is $y(x_\text{max}) = 0$ at an undetermined $x_\text{max}$, and $y'(0) = 0$.  From Sec.\ref{sec-kink} we learned to replace $y'(0)=0$ by $y'(0)=-\tan\theta_c$ instead. Despite that it is not smooth, it does eliminate the boundary variation and is indeed what we get from the Euler-Lagrange equation. A more formal argument is to write down $E[x(y)]$ instead, for which the standard choice, $x=0$ and $x'=0$ does not exclude the kink.  The resulting Euler-Lagrange  equation is essentially the same and we can just use it. 

We will keep it simple and solve the Euler-Lagrange equation for $y(x)$.
\begin{equation}
\Delta V = \frac{d}{dx} \left[\frac{d\sigma}{dy'}\sqrt{1+y'^2}+\sigma\frac{y'}{\sqrt{1+y'^2}}\right]
=\frac{d}{dx}\left(\sigma\sin\theta+\frac{d\sigma}{d\theta}\cos\theta\right)~.
\end{equation}
The general solution can be parametrized by $\theta$,
\begin{equation}
x(\theta) = const. + \frac{1}{\Delta V}
\left(\sigma\sin\theta+\frac{d\sigma}{d\theta}\cos\theta\right)~.
\label{eq-x}
\end{equation}
Note that the quantity in the parenthesis is zero at both $\theta = \theta_c$ and $\theta=0$.  Therefore, solutions starting at either value will eliminate boundary variations as promised, and also set that integration constant to zero. We can then integrate to find
\begin{equation}
y(\theta) = \frac{1}{\Delta V}
\left(\sigma\cos\theta-\frac{d\sigma}{d\theta}\sin\theta\right)~.
\end{equation}

We can try to generalize this to $N$ dimensions, such as taking $x_N$ as the longitudinal direction along $(\vec{\phi}_+-\vec{\phi}_-)$ which we will denote by $x_L$, and $x_1$ to $x_{N-1}$ as the transverse directions.  Be aware that in general the structure of the potential can further break the $SO(N-1)$ symmetry, since the exact interpolation between $\vec\phi_{\pm}$ may still involve nontrivial profile of the transverse fields.  That being said, we will focus on the simple cases with $SO(N-1)$ symmetry, in which $\sigma$ is again only a function of $\theta$.  We can simply write down the energy
\begin{equation}
E[x_L(x_T)] = 2 S_{N-2} \int  dx_T \,x_T^{N-2}
\left(\sigma(\tan^{-1}x'_L)\sqrt{1+x_L'^2}-\Delta V x_L \right)~,
\end{equation}
where $S_{N-2}$ is the area for an $(N-2)$ unit sphere, $x_T=\sqrt{\sum_{i=1}^{N-1} x_i^2}$.

This leads to the general solution,
\begin{eqnarray}
x_T(\theta) &=& \frac{(N-1)}{\Delta V}
\left(\sigma\sin\theta+\frac{d\sigma}{d\theta}\cos\theta\right)~, \\
x_L(\theta) &=& \frac{(N-1)}{\Delta V}
\left(\sigma\cos\theta-\frac{d\sigma}{d\theta}\sin\theta\right)~.
\label{eq-xN} 
\end{eqnarray}

\begin{figure}
\begin{center}
\includegraphics[width=2.5in]{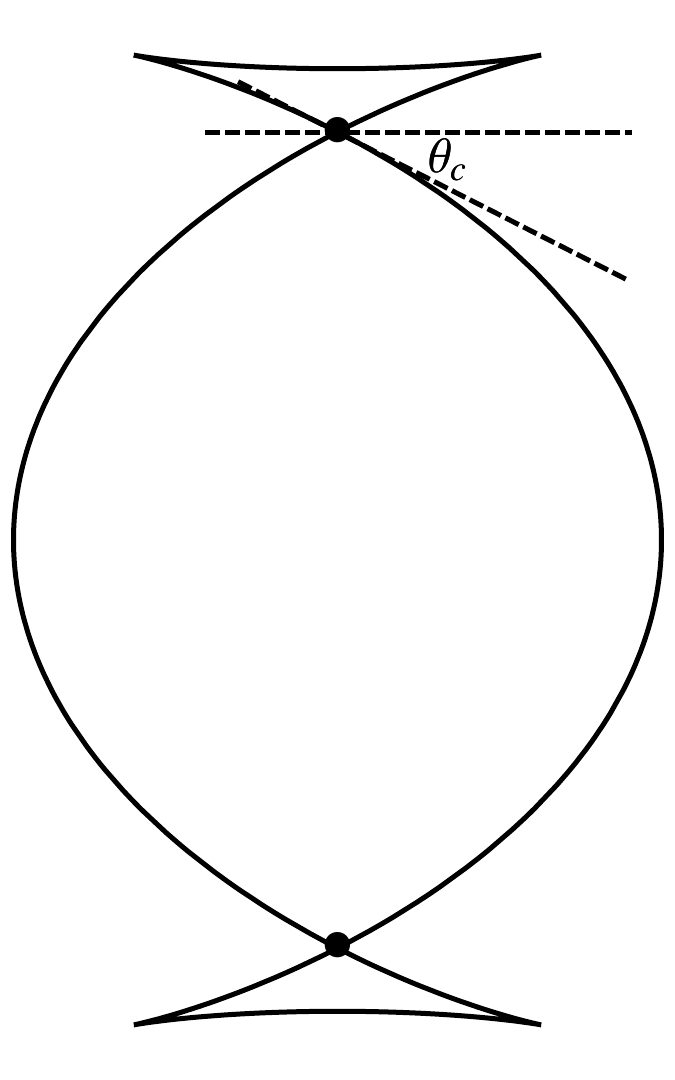} 
\end{center}
\caption{The bubble profile given by Eq.~(\protect\ref{eq-xN}) and (\protect\ref{eq-sigma}) when $c_L>2c_T$.  The correct profile of the critical bubble is simply the middle portion.
}
\label{fig-candy}
\end{figure}

Although this is a na\"ive generalization of \cite{RudBru95}, we should take a closer look.  Note that by symmetry, we have $\frac{d\sigma}{d\theta}=0$ at $0$ and $\pi/2$.  Also for simplicity we can treat $\sigma(\theta)$ as a monotonically decreasing function.  So we can see that $x_N$ is positive definite, but there is a risk of  $x$ being negative.  Since $x(\pi/2)$ is still always positive definite, and $x$ goes to zero exactly at $\theta_c$, if we na\"ively plot Eq.~(\ref{eq-xN}) we may get something like a wrapped candy, as in Fig.\ref{fig-candy}. 

We will provide a simple argument to prove the following statement.

{\bf Eq.~(\ref{eq-xN}) always gives the correct critical bubble profile.  When the flat longitudinal domain-wall is stable, it works with $\boldsymbol{0 < \theta<\pi/2}$.  When the flat longitudinal domain-wall is unstable, we should take the largest $\boldsymbol{\theta_c}$ such that $\boldsymbol{x_T(\theta_c)=0}$ and use the portion $\boldsymbol{\theta_c<\theta<\pi/2}$.}  In other words, cut the extra wrappings and keep the candy.

First of all, the saddle point we are looking for has only one negative mode, which corresponds to the  expansion/contraction of the bubble.  This means that fluctuations of the wall shape should still correspond to positive modes.  So locally every wall segment still settles to the minimum energy configuration.  As shown in Sec.\ref{sec-kink}, domain-walls with $\theta>\theta_c$ can stay, but those with $\theta<\theta_c$ cannot exist on the critical bubble profile.  Therefore, we have to cut off the tails, and the profile necessarily includes a kink.

Next, can the kink occur at some $\theta>\theta_c$?  Picture this in 2D for better intuition, that is like using a smaller portion of the two shells.  Also, can we take the two shells further apart and interpolate between them with zigzag walls?  The former possibility is making the bubble smaller, while the later is making it bigger.  Through a pictorial argument, we can show that they both make the total energy smaller, establishing that Eq.~(\ref{eq-soln}) is really the saddle point with this unique negative mode.  The foundation of our argument is
\begin{equation}
\Delta V x_L(\theta_c) =(N-1) \frac{\sigma(\theta_c)}{\cos\theta_c}~,
\label{eq-crit}
\end{equation}
which we can get from Eq.~(\ref{eq-xN}).  The physical meaning is that the energy difference due to volume for a cylinder---an $(N-1)$ sphere times height $x_L(\theta_c)$, is equal to the energy in the domain-wall that covers the $(N-1)$ sphere by a zigzag profile with orientation $\theta_c$.

\begin{figure}
\begin{center}
\includegraphics[width=10cm]{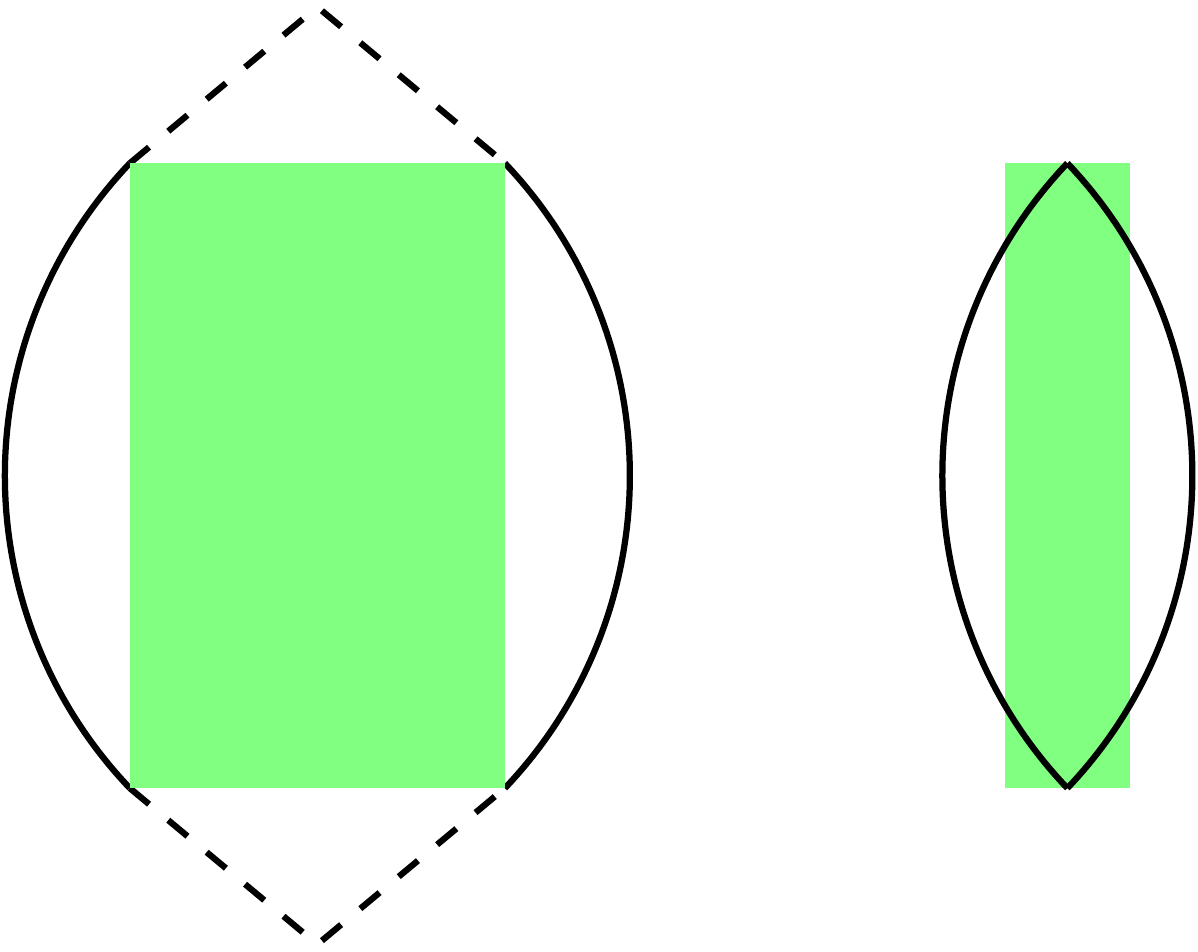}
\caption{The left figure visualizes Eq.~(\protect\ref{eq-insert}), where we attempt to make a larger bubble by inserting true vacuum regions and extra interpolation walls.  The right figure visualizes Eq.~(\protect\ref{eq-remove}), where we try to make a smaller bubble by removing some part of the walls and the true vacuum region.  Both result in smaller total energy, which shows that the kinky shape is indeed a saddle point.
\label{fig-saddle}}
\end{center}
\end{figure}

Then, as shown in the left portion of Fig.\ref{fig-saddle}, the energy lost due to the green (shaded) region is equal to the contribution from the dotted domain-wall.  
After they cancel each other, the two extra triangular regions still contribute $-\Delta V$, so the total energy is indeed less. 
\begin{equation}
({\rm zigzag \ wall})=\Delta V ({\rm rectangle})
<\Delta V (\rm extra \ false \ vacuum \ region)~.
\label{eq-insert}
\end{equation}

In the right portion of Fig.\ref{fig-saddle} we try to make a smaller bubble by removing the true vacuum region and domain-walls covered by the green (shaded) rectangle.  Then patching the remaining two shells together as a smaller bubble, with a kink angle larger than $\theta_c$.  Since $\sigma(\theta_c)/\cos\theta_c$ is a minimum of $\sigma(\theta)/\cos\theta$, losing those wall segments over-compensates the energy gain even if we remove $-\Delta V$ of the entire green (shaded) rectangle, and there are even those 4 corners that we are not really removing.  So the energy of the resulting smaller bubble is also less.
\begin{equation}
({\rm removed \ wall})> ({\rm zigzag \ wall})=
\Delta V ({\rm rectangle})
>\Delta V (\rm removed \ false \ vacuum \ region)~.
\label{eq-remove}
\end{equation}

\subsection{Smooth Bubbles}

We can get more intuition by solving the exact bubble shape from a specific $\sigma(\theta)$.  In the first example we will use Eq.~(\ref{eq-sigmamotion}), where no spontaneous symmetry breaking should occur.  Thus, we are expecting a smooth bubble.  Plugging into Eq.~(\ref{eq-xN}), we get
\begin{eqnarray}
x_T(\theta) &=& \frac{N-1}{\Delta V} \frac{\sigma(\pi/2)^2\sin\theta}
{\sqrt{\sigma(0)^2\cos^2\theta + \sigma(\pi/2)^2\sin^2\theta}}~, \\
x_L(\theta) &=& \frac{N-1}{\Delta V} \frac{\sigma(0)^2\cos\theta}
{\sqrt{\sigma(0)^2\cos^2\theta + \sigma(\pi/2)^2\sin^2\theta}}~.
\end{eqnarray}

Obviously, the bubble takes the shape of an ellipsoid,
\begin{equation}
\frac{\sum_{i=1}^{N-1}x_i^2}{c_T^2}+\frac{x_L^2}{c_L^2} = (N-1)^2 r_0^2~,
\end{equation}
where 
\begin{equation}
r_0 = \frac{\sigma(0)}{c_L \,  \Delta V} = \frac{\sigma(\pi/2)}{c_T\,  \Delta V} 
= \frac{1}{\Delta V} 
\int_{\rm path} \sqrt{2V} |d\vec\phi|
\end{equation}
comes from Eq.~(\ref{eq-simpletension}).

It is then straightforward to calculate the saddle point energy,
\begin{equation}
E_s = S_{N-1}\frac{(N-1)^{N-1}}{N} 
\frac{\left(\int_{\rm path} \sqrt{2V} |d\vec\phi|\right)^N}{\Delta V^{N-1}}
c_L c_T^{N-1}~.
\end{equation}
Compare this answer to the usual form people use assuming a spherical bubble, Eq.~(\ref{eq-rate}), the difference can be characterized by an effective Fermi velocity, 
\begin{equation}
v_F \rightarrow (c_L c_T^{N-1})^{1/N}~,
\label{eq-Fermi}
\end{equation}
which is a weighted geometric average of sound speeds.

\subsection{Kinky Bubbles}

Now we turn our attention to Eq.~(\ref{eq-sigma}).  Plugging  it into Eq.~(\ref{eq-xN}), we get
\begin{eqnarray}
x_T(\theta) &=& \frac{(N-1)}{\Delta V}
\left(\bigg[2\sigma(\pi/2)-\sigma(0)\bigg]\sin\theta+\bigg[\sigma(0)-\sigma(\pi/2)\bigg]\sin^3\theta\right)~, \\
x_L(\theta) &=&  \frac{(N-1)}{\Delta V}
\left(\bigg[2\sigma(0)-\sigma(\pi/2)\bigg]\cos\theta-\bigg[\sigma(0)-\sigma(\pi/2)\bigg]\cos^3\theta\right)~.
\label{eq-soln}
\end{eqnarray}
As expected, when $\sigma(0)<2\sigma(\pi/2)$, we still have a smooth bubble profile.  When $\sigma(0)>2\sigma(\pi/2)$, as proved in Sec.\ref{sec-shape} we just use the portion $\theta_c<\theta<\pi/2$.

The expression of $E_s$ is quite complicated in arbitrary dimensions, so we only present the ``realistic'' dimensions.  For $N=2$, we have
\begin{eqnarray}
E^{N=2}_s &=& 
\frac{\left(\int_{\rm path}\sqrt{2V}|d\vec\phi|\right)^2}{4\Delta V}
(10c_Lc_T-c_L^2-c_T^2) \frac{\pi}{2}~, \ \ \ {\rm for} \ \ \ c_L<2c_T~, \\
&=& \frac{\left(\int_{\rm path}\sqrt{2V}|d\vec\phi|\right)^2}{4\Delta V}
\bigg( (10c_Lc_T-c_L^2-c_T^2) \cos^{-1}\sqrt{\frac{c_L-2c_T}{c_L-c_T}}
\nonumber \\ 
&+& (c_L+13c_T)\sqrt{(c_L-2c_T)c_T} \bigg)~, \ \ \ {\rm for} \ \ \ c_L>2c_T
\end{eqnarray}

This is quite complicated.  We should again compare it to the spherical bubble and think in terms of the effective Fermi velocity, especially in the limit $c_L\gg c_T$.
\begin{eqnarray}
v_F &\rightarrow& \left(\frac{10c_Lc_T-c_L^2-c_T^2}{8}\right)^{1/2}~,
\ \ \ {\rm for} \ \ \ c_L<2c_T~, \nonumber \\
v_F &\rightarrow& \frac{4}{\sqrt{3}}(c_Lc_T^3)^{1/4}~, 
\ \ \ {\rm for} \ \ \ c_L\gg c_T~.
\end{eqnarray}

For $N=3$, we have
\begin{eqnarray}
E^{N=3}_s &=& 4\pi 
\frac{\left(\int_{\rm path}\sqrt{2V}|d\vec\phi|\right)^3}{\Delta V^2}
\frac{4(c_L^2-10c_L^2c_T+52c_Lc_T^2-8c_T^3)}{105}~,
\ \ \ {\rm for} \ \ \ c_L<2c_T~, \nonumber \\
&=& 4\pi 
\frac{\left(\int_{\rm path}\sqrt{2V}|d\vec\phi|\right)^3}{\Delta V^2}
\frac{32c_T^2(7c_L-6c_T)\sqrt{c_T}}{105\sqrt{c_L-c_T}}~,
\ \ \ {\rm for} \ \ \ c_L>2c_T~.
\end{eqnarray}
Similarly we have
\begin{eqnarray}
v_F \rightarrow 
\left(\frac{8}{5}c_L^{1/2}c_T^{5/2}\right)^{1/3}~, 
\ \ \ {\rm for} \ \ \ c_L\gg c_T~.
\end{eqnarray}

Comparing these to Eq.~(\ref{eq-Fermi}), we found that the effective Fermi velocity is still a weighted geometric mean, but the weight on $c_L$ is always reduced by half.  This is quite understandable since the critical bubble approaches a thin slit.  A major portion of its domain-wall is aligned in the transverse direction.  It is straightforward to show that this limit generalizes to $N$ dimensions as
\begin{equation}
v_F \sim (c_L^{1/2}c_T^{N-1/2})^{1/N}~.
\end{equation}

\section{Conclusion}
\label{sec-dis}

We studied the orientation dependence of the domain-wall tension in a vector field theory.  We then constructed critical bubbles for thermal nucleation.  The shape of the bubble directly depends on the tension through a simple formula, Eq.~(\ref{eq-xN}).  The longitudinally oriented domain-wall is usually the most massive, thus it may spontaneously break into zigzag segments of a critical orientation $\theta_c$.  When that happens, the critical bubble develops two kinks of angle $\theta_c$, and the overall shape is still described by Eq.~(\ref{eq-xN}) with a careful interpretation.\footnote{Analysis of the bubble shape for nucleation is identical to those of equilibrium bubbles, known as the Wulff construction.  Earlier works\cite{GalFou95,Leg75,Whe75,He3,Fou95,MacJia95,RudLoh99,SilPat06} have qualitatively similar results.  We further specify that zigzag segments of $\theta_c$ is the configuration to which the instability settles.  The recognition and interpretation of the kinky bubble shape is also more transparent in our analysis.  The tension previously studied is often expanded as $\sigma(\theta) = \sigma_0 + a\cos\theta + b\cos 2\theta$, and most analysis focused on the effect of $a\neq0$.  In our model there is a reflection symmetry---the domain-wall tension does not change when you look at it from the other side.  Thus we always have $a=0$.  This makes our situation closer to a 2D lattice model\cite{AleSur90}, where similar bubble shape was observed.}

Our analytic and numerical study shows that the freedom to take different paths in a multi-dimensional field space is essential for the instability.  If we choose parameters that reduce the number of dynamical fields down to one, the longitudinal wall is always stable.  That is however an extreme choice.  For typical choices of parameters, at $c_L/c_T>2$ the longitudinal wall becomes unstable, and the critical bubble develops two kinks.  Such behavior can appear with an even smaller sound speed ratio, $c_L/c_T>\sqrt{2}$, if we tune the potential to the other extreme limit.  This range of sound speed ratio is not hard to find in real materials.

We pick two representative forms of $\sigma(\theta)$, given by Eq.~(\ref{eq-sigma}) and (\ref{eq-sigmamotion}), to calculate the exact shapes of critical bubbles.  This allows us to observe the scaling property of tunneling rates.  When the critical bubble is deformed but still smooth, we can modify the standard tunneling rate formula, Eq.~(\ref{eq-rate}), through an effective Fermi velocity,
\begin{equation}
v_F \rightarrow (c_L c_T^{(N-1)})^{1/N}~.
\end{equation}
This is quite intuitive since one particular orientation is longitudinal, and all others are transverse.  They care about the sound speeds in their own orientations.  When the bubble starts to develop kinks and we further increase the sound speed ratio, the scaling behavior changes to
\begin{equation}
v_F \rightarrow (c_L^{1/2} c_T^{(N-1/2)})^{1/N}~.
\end{equation}
This is because the kink-development removes a large portion of the longitudinally oriented wall, so $c_L$ becomes less important.

On top of modifying the tunneling rate estimation, our result has a practical impact.  Typically, the experimental measurement of domain-wall tension involves measuring the bubble radius\cite{OshCro77}.  That is done by observing a domain-wall popping through a partition with holes.  When it does, the radius of the hole is identified with the bubble radius.  Our result shows that for vector fields, the orientation of that partition is important.  Only for a longitudinally oriented partition, the popping radius can be identified with $x$ given by Eq.~(\ref{eq-xN}).  For other orientations, the hole and the bubble do not have common symmetries.  Therefore the exact relation between the popping radius and the critical radius requires further analysis.

We have only taken a small step toward a rich phenomenology.  Given the new insight here, many nontrivial questions arise.  How does the domain-wall move/bubble expand given this orientation dependence? Especially when there is a kink, can we expect the tip to travel at $c_L$, leaving behind a Cherenkov-like tail of domain-walls bounded by $c_T$?  How does the spontaneously broken planar symmetry interact with impurities or other external effects?  All these await future study, and may lead towards a more practical understanding about some exotic theories of phase transitions which relies on the properties of domain-walls\cite{EasGib09,GibLam10,YanTye11}.

\acknowledgments 
We thank Igor Aleiner, Solomon Endlich, and Erick Weinberg for stimulating discussions.  This work is supported in part by the US Department of Energy, grant number DE-FG02-11ER41743.

\bibliographystyle{utcaps}
\bibliography{all}

\appendix

\section{Examples of different potentials.}
\label{sec-examples}

Here we present some analytical approaches and the numerical evaluation of the domain-wall tension given by the relaxation method\cite{AguJoh09a,GibLam10,AhlGre10}.  The first potential we study is a double well in one direction and a quadratic in the other direction.
\begin{equation} \label{eq-DoubleWell}
V(\phi_x, \phi_y)=  - \frac{1}{2}\mu^2  \phi_y^2 + \frac{1}{4} \lambda \phi_y^4 + \frac{1}{2} \beta \phi_x^2~.
\end{equation}
This is qualitatively similar to Eq.~(\ref{eq-potmot}) when $m^2>0$.  The differences are some 4th order terms involving $\phi_x$, which is not very important when $\beta>0$ stabilizes a trajectory near $\phi_x=0$.  The two potentials can be roughly related by
\begin{eqnarray}
-\frac{\mu^2}{2} &=& \frac{m^2}{2}+b|\vec{H}|^2~, \nonumber \\
\frac{\beta^2}{2} &=& \frac{m^2}{2}~.
\end{eqnarray}
The two degenerate minima sit at $(0,\pm\sqrt{\frac{\mu^2}{\lambda}})$.
For the purely longitudinal (or transverse) wall oriented along the $\vec{y}$ (or $\vec{x}$), we can solve the problem analytically and get the exact value of the tension. 

Longitudinal wall ($\theta = 0$):
\begin{align}
	& \phi_y(x,y) = \sqrt{\frac{\mu}{\lambda}} \tanh \left(\frac{\mu y}{c_L \sqrt{2}}\right) \\
	& \phi_x(x,y) = 0 \\
	& \sigma = \sigma(0) = \frac{2 \sqrt{2} \mu^3 c_L}{3 \lambda}~.
\end{align}
Transverse wall ($\theta = \frac{\pi}{2}$):
\begin{align}
	& \phi_y(x,y) = \sqrt{\frac{\mu}{\lambda}} \tanh \left(\frac{\mu x}{c_T \sqrt{2}}\right) \\
	& \phi_x(x,y) = 0 \\
	&\sigma = \sigma(\pi/2) = \frac{2 \sqrt{2} \mu^3 c_T}{3 \lambda}~.
\end{align}

For other orientations of the wall, we can evaluate the tension numerically.  
Before that, we can analyze two extreme cases.  Using the method of rotating the potential as described in Sec.\ref{sec-orient}, we have
\begin{eqnarray}
\sigma(\theta) &=& 
\int dx ~ \frac{c_T^2}{2}\phi_x'^2 + \frac{c_L^2}{2}\phi_y'^2
+V_\theta(\phi_x,\phi_y) \nonumber \\
&=& \int dx ~ \frac{c_T^2}{2}\phi_x'^2 + \frac{c_L^2}{2}\phi_y'^2
+V(\phi_x\cos\theta+\phi_y\sin\theta,\phi_y\cos\theta-\phi_x\sin\theta)~.
\end{eqnarray}
When $\beta\rightarrow\infty$, we effectively have a single field problem with
\begin{equation}
\bar\phi = \frac{\phi_x}{\cos\theta} = \frac{\phi_y}{\sin\theta}~,
\end{equation}
such that
\begin{equation}
\sigma(\theta) = 
\int dx ~ \frac{c_L^2\cos^2\theta + c_T^2\sin^2\theta}{2}\bar\phi'^2
+V(\bar\phi,0)~.
\end{equation}
Clearly, this gives us Eq.~(\ref{eq-sigmamotion}).

\begin{figure}
   \centering
   \includegraphics[width=3in]{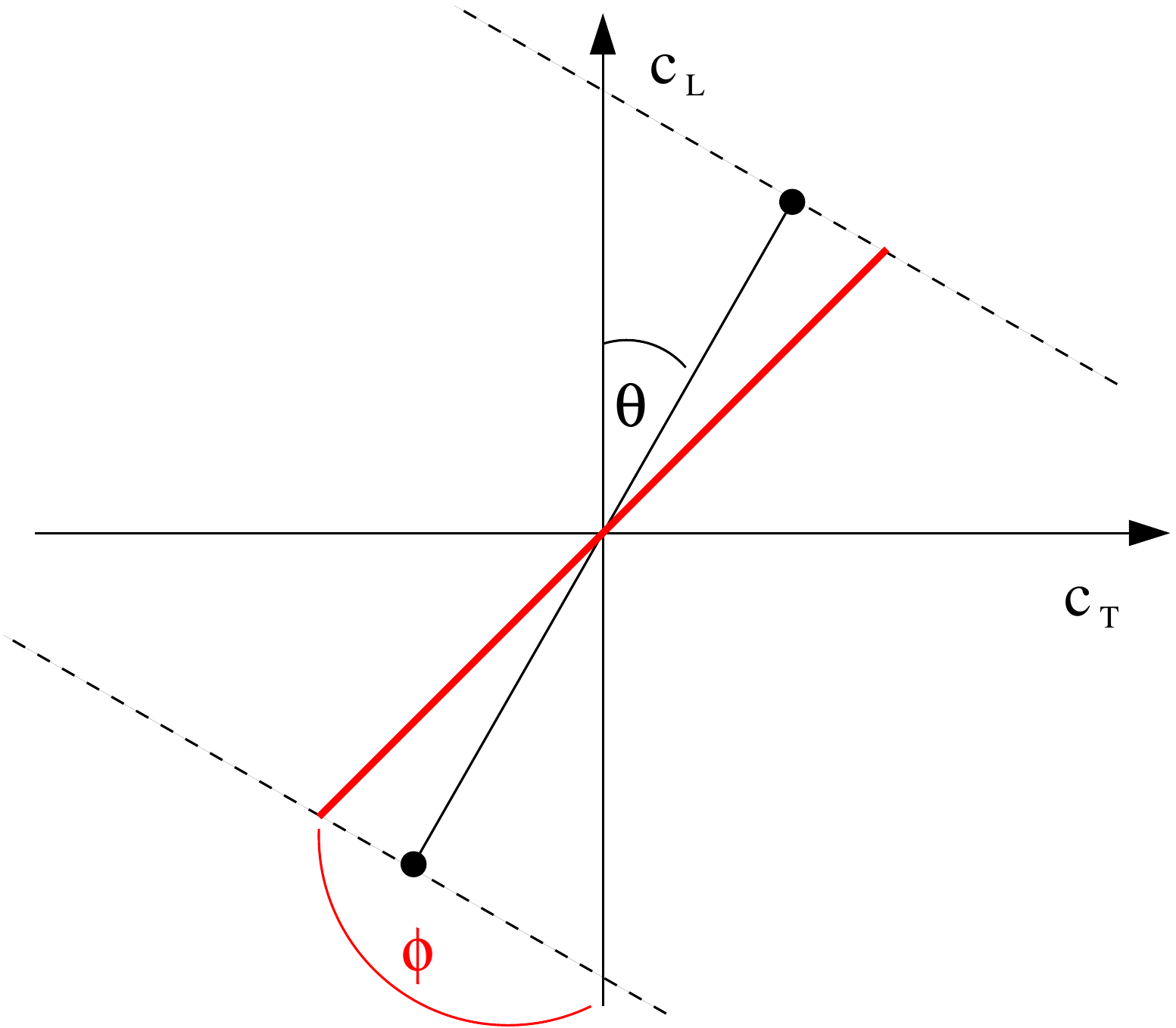} 
   \caption{The $c_L$ and $c_T$ axes are the directions in which the field has purely longitudinal and transverse sound speeds.  The two dots represent the two discrete vacua.  In the limit $\beta\rightarrow0$, the dashed lines through them are almost in the vacuum, too.  The important portion of the domain-wall is the red (thick) path from one line to the other, which is free to pick the best orientation $\phi$.}
   \label{fig-path}
\end{figure}

The other extreme limit is $\beta\rightarrow0$, at which the potential is flat in the $\phi_x$ direction.  The two degenerate vacua approach two separated lines.  Moving along these lines contributes nothing to the tension.  As shown in Fig.~\ref{fig-path}, the path that minimizes the tension involves first moving along these lines to an appropriate angle $\phi$, then connecting through a straight line.  The tension of this path is a function of both $\theta$ and $\phi$ through the orientation dependence in Eq.~(\ref{eq-sigmamotion}), and a simple projection of the length.
\begin{equation}
\sigma(\theta,\phi) = \frac{\sigma(0)}{c_L}\frac{1}{\cos(\theta-\phi)}
\sqrt{c_L^2\cos^2\phi + c_T^2\sin^2\phi}~.
\end{equation}
Minimizing this with $\phi$, we have
\begin{equation}
\phi_m(\theta) = \arccos \frac{c_T^2\cos\theta}
{\sqrt{c_T^4\cos^2\theta+c_L^4\sin^2\theta}}~.
\end{equation}
So the tension in this case should be
\begin{equation}
\sigma(\theta) = \sigma[\theta,\phi_m(\theta)]~.
\label{eq-sigmaextreme}
\end{equation}

We can apply the analysis in Sec.\ref{sec-kink} and calculate the stability condition for the flat longitudinal wall.
\begin{equation}
\frac{1}{\sigma(0)}\frac{d^2\sigma}{d\theta^2}
=\left(1-\frac{c_L^2}{c_T^2}\right)>-1~.
\end{equation}
We can see that the wall becomes unstable as soon as $c_L>\sqrt{2}c_T$.

\begin{figure}
   \centering
   \includegraphics[width=3in]{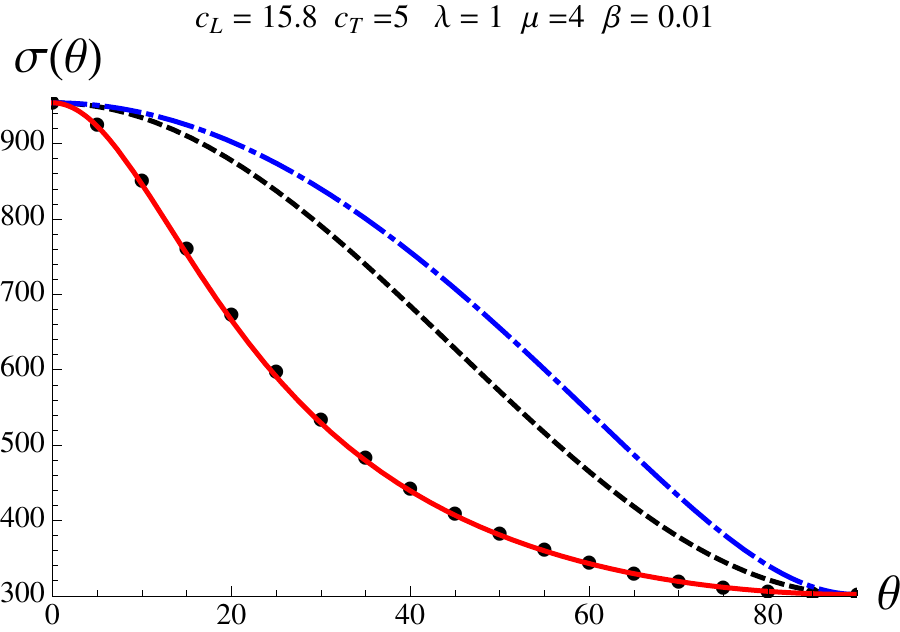} 
   \includegraphics[width=3in]{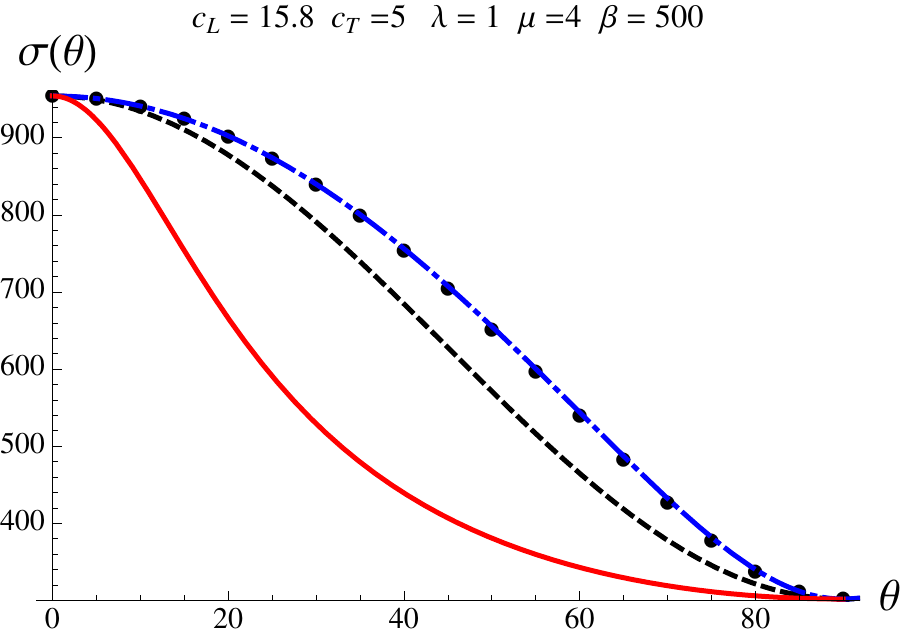} 
   \caption{The numerically calculated values of the tension for a double-well potential are shown in dots.  The three analytical fits: Eq~.(\protect\ref{eq-sigma}) is the dashed line, Eq.~(\protect\ref{eq-sigmamotion}) is the dot-dashed (blue) line, and Eq.~(\protect\ref{eq-sigmaextreme}) is the solid (red) line.  We can see that in the for $\beta$, Eq.~(\protect\ref{eq-sigmamotion}) is a good fit, and for small $\beta$, Eq.~(\protect\ref{eq-sigmaextreme}) is a good fit.}
   \label{fig-extreme}
\end{figure}

\begin{figure}
   \centering
   \includegraphics[width=3in]{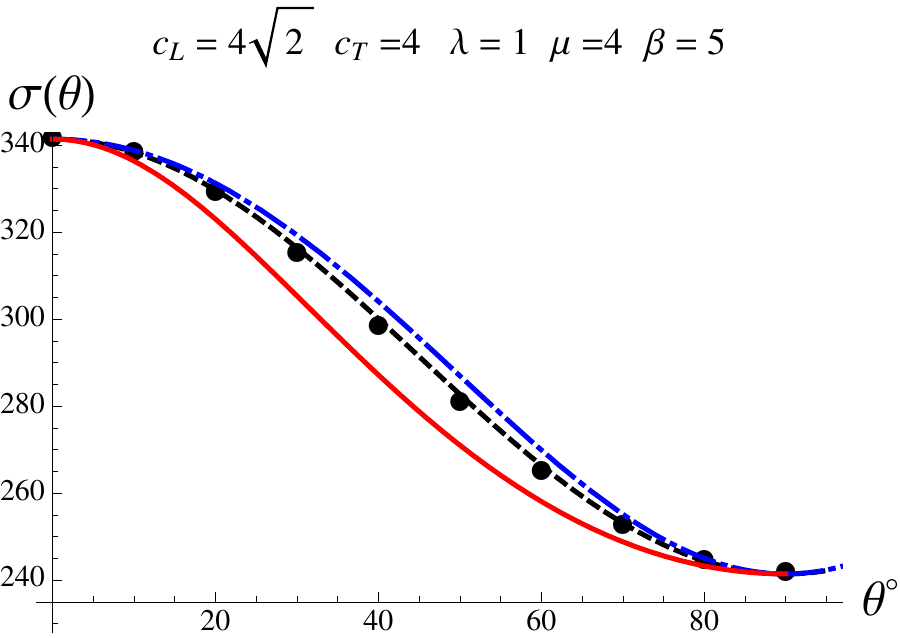} 
   \includegraphics[width=3in]{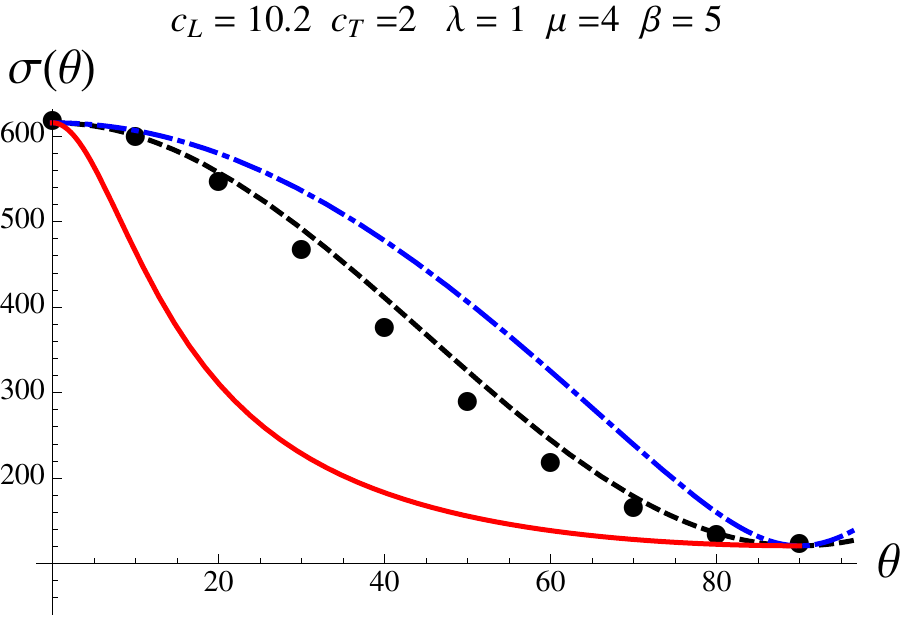} 
   \caption{The numerically calculated values of the tension for a double-well potential are shown in dots.  Again the three analytical fits: Eq~.(\protect\ref{eq-sigma}) is the dashed line, Eq.~(\protect\ref{eq-sigmamotion}) is the dot-dashed (blue) line, and Eq.~(\protect\ref{eq-sigmaextreme}) is the solid (red) line.  The two figures use the same potential but different sound speed ratios.  }
   \label{DoubleWellNumericalTension}
\end{figure}

We next provide several plots with the numerical values on top of the three possible fits, Eq.~(\ref{eq-sigma}), (\ref{eq-sigmamotion}) and (\ref{eq-sigmaextreme}).  Fig.~\ref{fig-extreme} shows that the two extreme limits indeed fit very well with our analysis.  Fig~.(\ref{DoubleWellNumericalTension}) shows that with a more moderate choice of parameters, Eq.~(\ref{eq-sigma}) is quite reliable independent in various sound speeds.

\begin{figure}
   \centering
   \includegraphics[width=10cm]{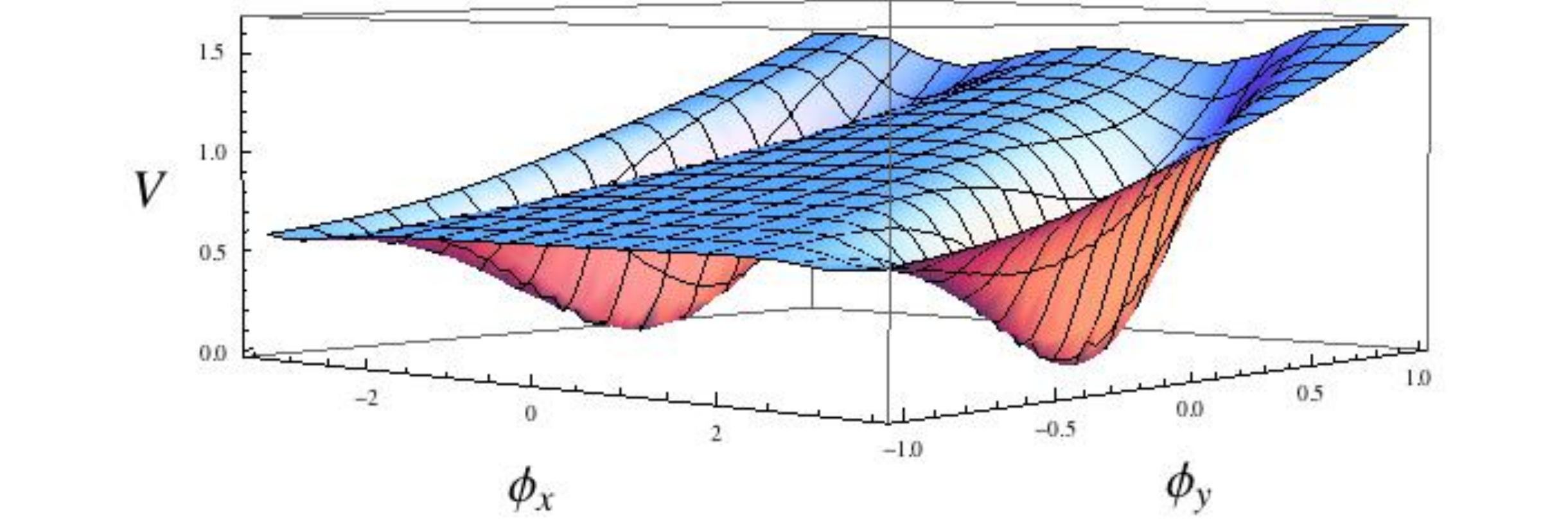} 
   \caption{The more complicated potential introduced in Eq~.(\protect\ref{ToyPotential}), with $q=0.5~, \ r_1 = -2.5~, \ r_2=2.5~.$  We will use two values of $S$, $1.1$ and $0.9$, but that makes no visual difference.}
   \label{WeirdPotential}
\end{figure}

In the end, we provide a much more complicated potential as in Fig~.(\ref{WeirdPotential}).  It has a general slope in the $\phi_x$ direction and two minima located at $(0.001,\pm2.498)$.  So trivially, the interpolation path will always involve both fields.
\begin{equation}\label {ToyPotential}
	V(\phi_x,\phi_y) = e^{q \phi_x} \left\{1 - S \exp\left[ -4 \left(\phi_x - \sin (\frac{\phi_y-r_1}{r_2-r_1})\right)^2\right] \right\} \left[ \tanh^2\left( \frac{(\phi_y-r_1)(\phi_y-r_2)}{3}\right)\right]~.
\end{equation}

\begin{figure}[htbp] 
   \centering
   \includegraphics[width=3in]{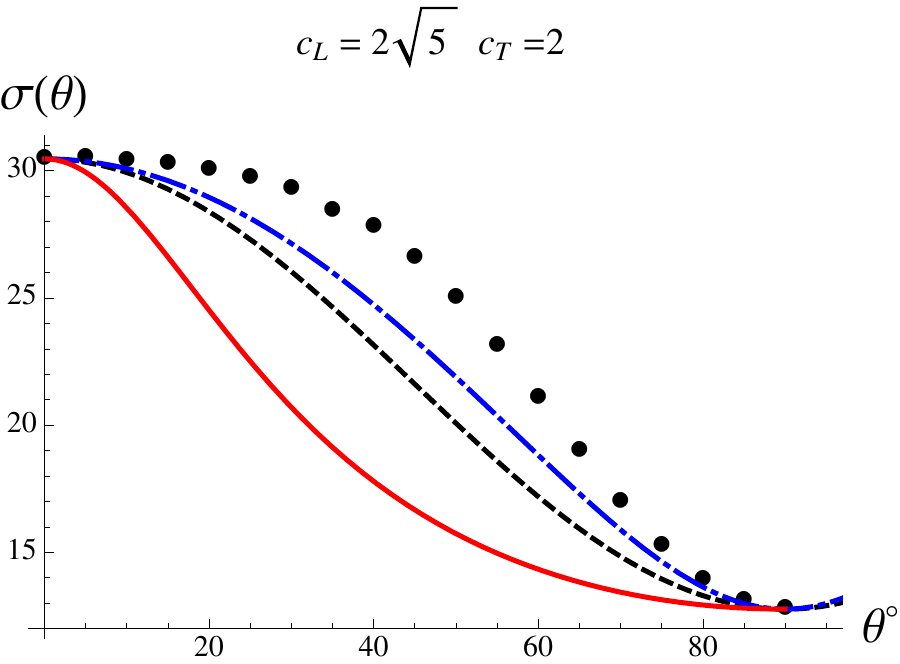} 
   \includegraphics[width=3in]{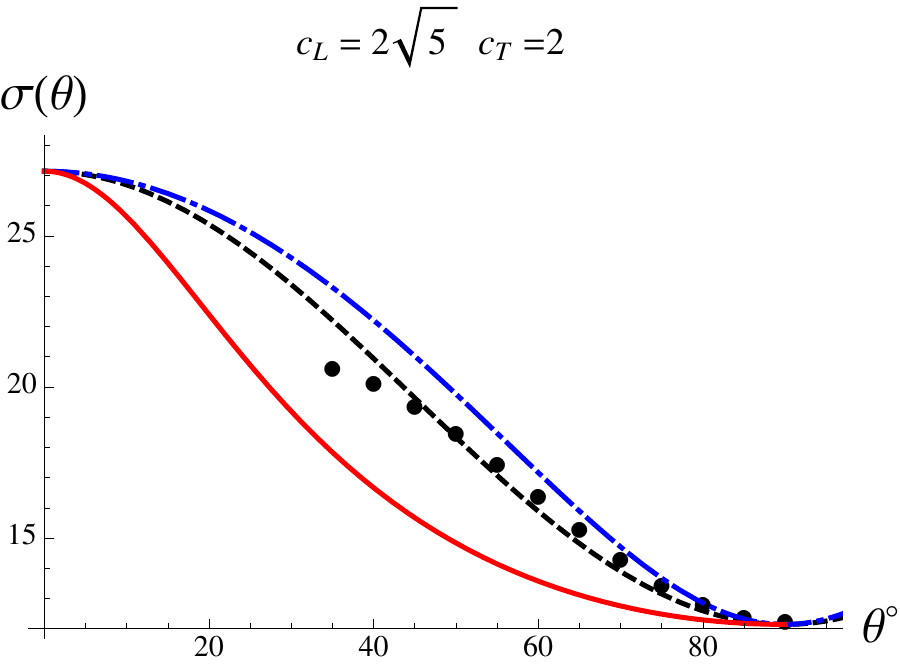}
   \caption{From the potential given by Eq~.(\protect\ref{ToyPotential}), we again compare the numerical $\sigma(\theta)$ with the three equations.  In the left figure we have $S=1.1$. In the right figure we have $S=1$ and for some orientations the domain-wall does not exist because the path runs away toward the $-\phi_x$ direction.}
   \label{fig:example}
\end{figure}

We numerically evaluated the tension for various orientations and plotted it against the three analytical fits in Fig.~(\ref{fig:example}).   The overall shape can be quite different from any equation given in this paper.  In particular, note that in the right portion of Fig.~(\ref{fig:example}) the longitudinal domain-wall (actually an open set near $\theta=0$) does not exist.\footnote{This comes from the same reason as described in \cite{AguJoh09a}.  The interpolation path breaks into two parts, connecting each vacuum individually with the $-\phi_x$ region.  For vector fields, such runaway behavior also acquires an orientation dependence.} 

\end{document}